\documentclass[12pt]{iopart}

\usepackage{iopams}  
\usepackage{graphicx}

\newcommand{\dz} {\partial_z}
\newcommand{\dzz} {\partial_{zz}}
\newcommand{\cF} {{\cal F}}

\newcommand{\eg} {{e.g., }}

\newcommand{\ie} {{i.e., }}

\newcommand{\rmB} {{\rm B}}

\newcommand{\rmL} {{\rm L}}

\newcommand{\rmT} {{\rm T}}
\newcommand{\vecf} {{\bf f}}
\newcommand{\vecq} {{\bf q}}
\newcommand{\vecr} {{\bf r}}
\newcommand{\vecv} {{\bf v}}

\newcommand{\xhat} {\hat{x}}
\newcommand{\yhat} {\hat{y}}

\begin{document}

\title 
{Hydrodynamic interaction in quasi-two-dimensional suspensions}

\author{H Diamant$^{1}$\footnote[1]{Author to whom correspondence 
should be addressed}, 
B Cui$^{2}$\footnote[2]{Present address: Department of Physics, 
Stanford University, Stanford, California 94305, USA},
B Lin$^2$ and S A Rice$^2$}

\address{$^1$ School of Chemistry, Raymond and Beverly Sackler Faculty
of Exact Sciences, Tel Aviv University,
Tel Aviv 69978, Israel}

\address{$^2$ Department of Chemistry, The James Franck Institute and
CARS, The University of Chicago, Chicago, Illinois 60637, USA}

\ead{hdiamant@tau.ac.il}

\begin{abstract}
  Confinement between two parallel surfaces is found, theoretically
  and experimentally, to drastically affect the hydrodynamic
  interaction between colloid particles, changing the sign of the
  coupling, its decay with distance and its concentration dependence.
  In particular, we show that three-body effects do not modify the
  coupling at large distances as would be expected from hydrodynamic
  screening.
\end{abstract}




\section{Introduction}
\label{sec_intro}

The dynamics of colloid suspensions and macromolecular solutions are
governed by hydrodynamic interactions, \ie correlations in the motions
of particles mediated by flows in the host liquid
\cite{colloids,DoiEdwards}. In various circumstances colloids are
spatially confined by rigid boundaries as in, \eg porous media,
biological constrictions, nozzles, or microfluidic devices.  Extensive
studies have been devoted recently to colloid dynamics at the
single-particle level, highlighting flow-mediated effects of the
boundaries \cite{1wall}--\cite{PRL04}. Confined suspensions have been
studied also by computer simulations \cite{Nagele}--\cite{Finland}.

Hydrodynamic interactions in an unconfined suspension \cite{Happel}
decay with inter-particle distance $r$ as $1/r$. They are positive,
\ie particles drag one another in the same direction. The long range
of the interaction leads to strong many-body effects, manifest, \eg in
an appreciable dependence of transport coefficients on particle volume
fraction. A particularly important many-body effect is hydrodynamic
screening \cite{DoiEdwards}---over length scales much larger than the
typical inter-particle distance the suspension responds to a slow
disturbance as if it were a homogeneous medium with merely an
increased viscosity.  This implies that at large distances many-body
effects change the prefactor of the $\sim 1/r$ pair
interaction.

\begin{figure}[tbh]
\centerline{\resizebox{0.3\textwidth}{!}
{\includegraphics{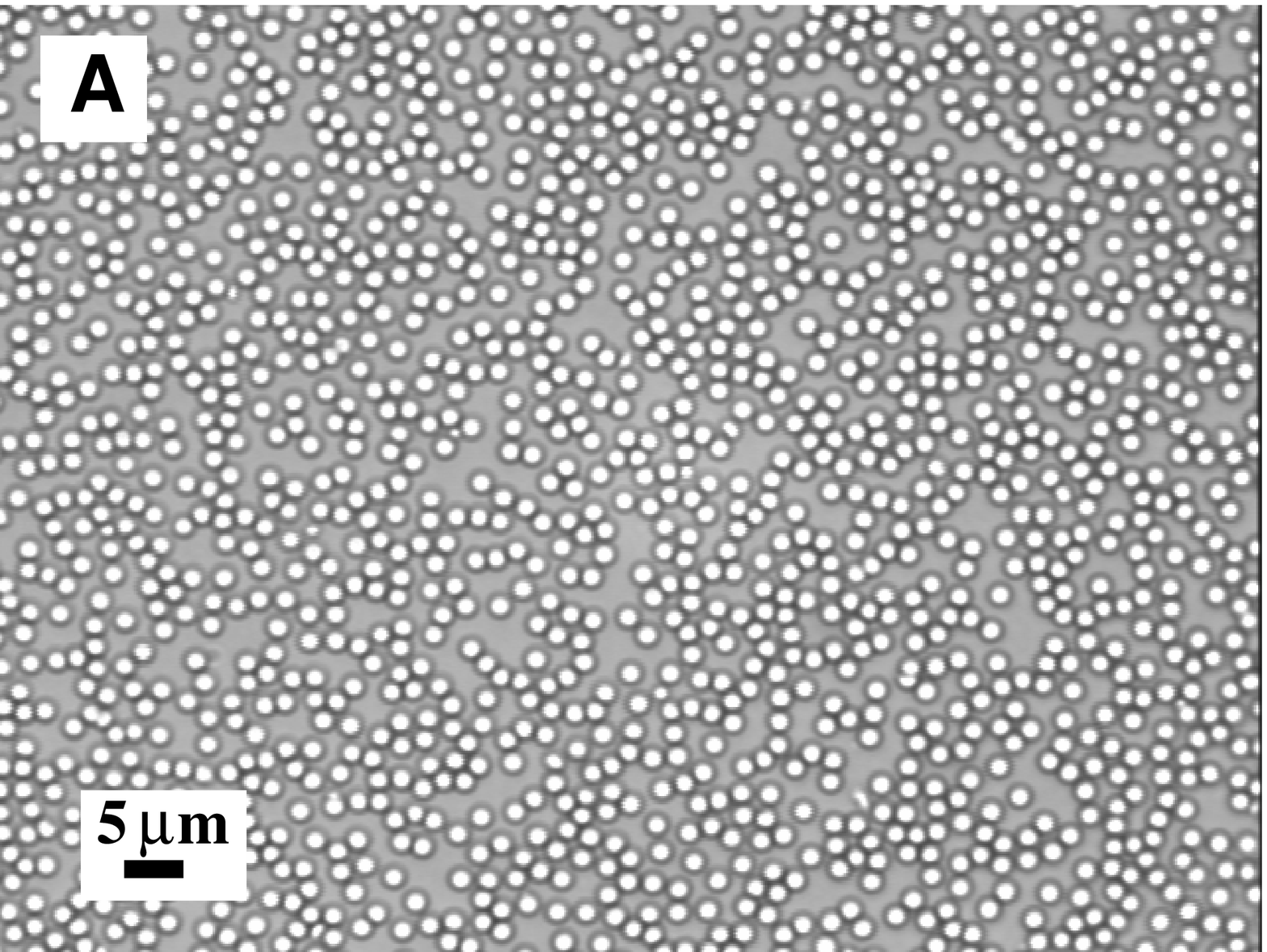}}
\hspace{0.2cm}
\resizebox{0.3\textwidth}{!}
{\includegraphics{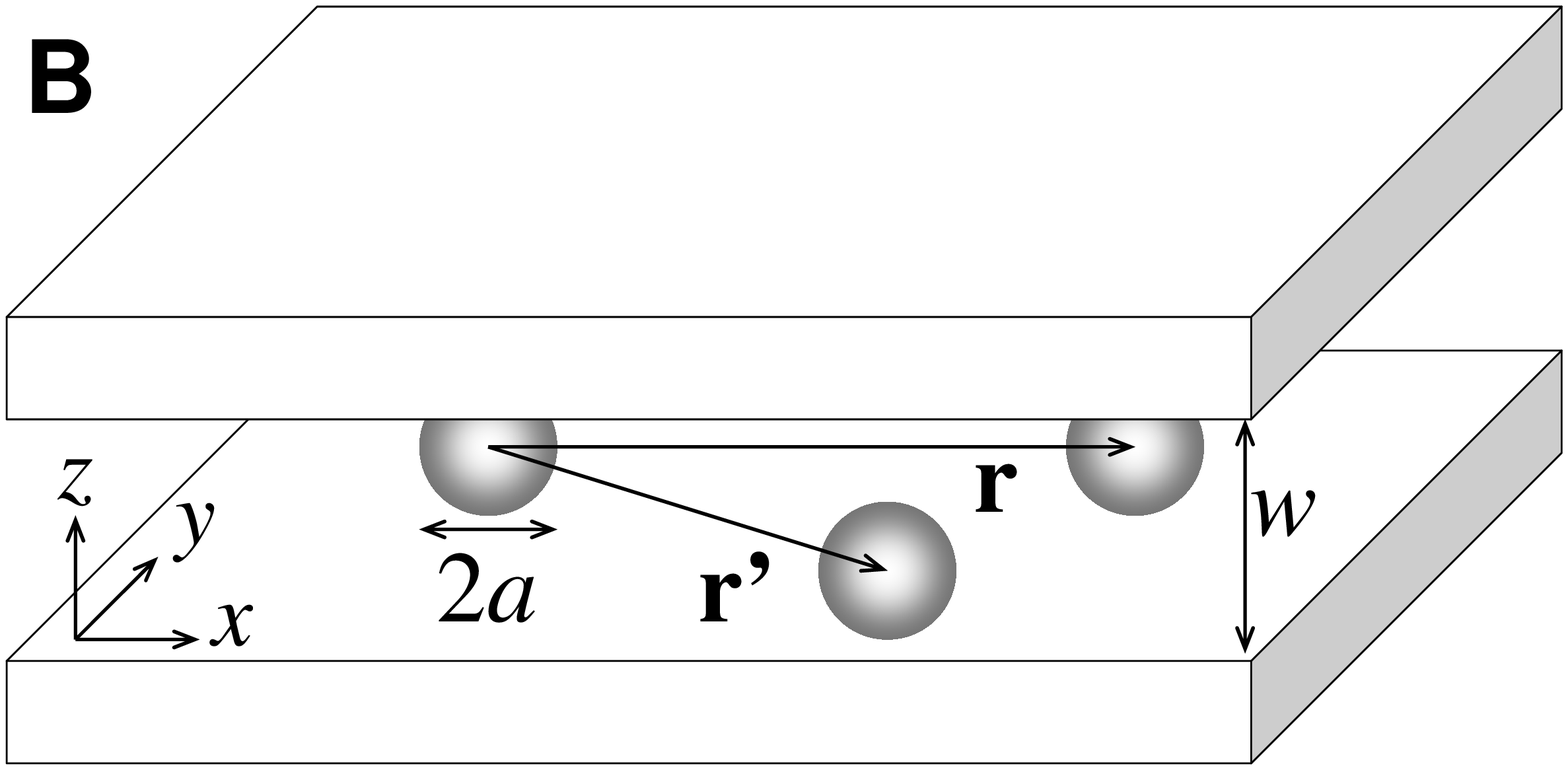}}}
\caption[]{($a$) Optical microscope image 
of the experimental Q2D suspension at area fraction $\phi=0.338$.
($b$) Schematic view of the system and its parameters.}
\label{fig_system}
\end{figure}

Our aim has been to investigate the effects of confinement on
hydrodynamic interactions. Elaborating on our previous short
publication \cite{PRL04}, we consider here the case of confinement
between two parallel plates, leading to a quasi-two-dimensional (Q2D)
suspension (figure \ref{fig_system}).  In section \ref{sec_def} we
define the problem and the corresponding nomenclature. In section
\ref{sec_pair} we address the hydrodynamic interaction between two
isolated particles, and in section \ref{sec_three} --- the three-body
correction at finite concentration. The experimental measurements are
presented in section \ref{sec_exp}, and in section \ref{sec_dis} we
discuss the results.

\section{Definition of the problem}
\label{sec_def}

The geometry considered in this work is depicted in figure
\ref{fig_system}($b$).  Identical, spherical particles of radius $a$
are suspended in a liquid of viscosity $\eta$ and temperature $T$,
confined in a slab of width $w$ between two planar solid surfaces. The
$x$ and $y$ axes are taken parallel, and the $z$ axis perpendicular,
to the surfaces. The particles are assumed to behave as hard spheres
with no additional equilibrium interaction. For simplicity we consider
cases where particle motion is restricted to two dimensions, \ie to a
monolayer at the mid-plane, $z=0$.  We use the notation
$\vecr(\brho,z)$ for three-dimensional position vectors, where
$\brho(x,y)$ is a two-dimensional position vector in the monolayer.
The area fraction occupied by particles is denoted by $\phi$.
The Reynolds number is assumed very low. (It is of order $10^{-6}$ in
the experimental system.)  The hydrodynamics,
therefore, are well described by viscous Stokes flows
\cite{Happel}. The confining boundaries are assumed impermeable and
rigid, imposing no-slip boundary conditions on the flow.

We characterize the pair hydrodynamic interaction between two
particles by the coupling mobilities $\Delta_{\rm L,T}(\rho)$ as
functions of the inter-particle distance $\rho$. These are the
off-diagonal terms in the mobility tensor of a particle pair, \ie the
proportionality coefficients relating the force acting on one particle
with the change in velocity of the other.  The two independent
coefficients, $\Delta_{\rmL}$ and $\Delta_{\rmT}$, correspond,
respectively, to the coupling along and transverse to the line
connecting the pair. [In reference \cite{PRL04} four coefficients were
considered, $\Delta_{\rm L,T}^\pm$, whose relation with the ones
considered here is $\Delta_{\rm L,T}=(\Delta_{\rm L,T}^+-\Delta_{\rm
L,T}^-)/2$.]

We use dimensionless quantities throughout this paper, scaling all
distances by the confinement length $w$ and all mobilities by
$B_0a/w$, where $B_0=(6\pi\eta a)^{-1}$ is the Stokes mobility of a
single, unconfined sphere. In places where we refer to dimensional
quantities for clarity, a tilde symbol is used, \eg $\tilde{r}=wr$.

\section{Pair interaction at infinite dilution}
\label{sec_pair}

We first examine the effect of confining boundaries on the flow due to
a local disturbance such as the one created by single-particle motion.
The flow field can be found by solving the Stokes and continuity
equations subject to the appropriate boundary conditions
\cite{Happel}. This was done for a point force (stokeslet) in a Q2D
geometry in reference \cite{Mochon}.  However, by returning to the
physical origin of the equations, one can gain a useful insight
applicable to various geometries and particle sizes.  The motion of a
particle perturbs the local liquid momentum and displaces liquid mass.
In an unconfined liquid the diffusion of the momentum disturbance,
whose leading moment is a momentum monopole, creates a far flow
decaying as $1/r$.  The mass term, whose leading moment must be a mass
dipole (since mass is neither created nor lost), adds to the far flow
a much smaller contribution $\sim 1/r^3$.  

In confinement, however, the situation is very different. It is the
momentum perturbation which, while diffusing away, is absorbed by the
boundaries and thus gives a contribution to the far field which is
exponentially small in $r=\tilde{r}/w$ in both Q2D and Q1D
geometries. The contribution from mass displacement propagates
laterally, creating a far flow parallel to the boundaries.  We
conclude that the far flow field induced by single-particle motion in
Q2D is that of a 2D mass dipole (source doublet).  If the particle
motion (mass dipole) is taken along the $x$ direction, we thus have a
far velocity field $\vecv(\vecr)$ of the form
\begin{equation}
  v_x(\vecr) \simeq f(z)(x^2-y^2)/\rho^4,\ \ \ 
  v_y(\vecr) \simeq f(z)(2xy)/\rho^4,\ \ \ 
  v_z(\vecr) \simeq 0.
\label{flowq2d}
\end{equation}
The 2D dipolar field must be modulated by a transverse profile $f(z)$
to make it vanish on the two confining surfaces. Note that the $xy$
dependence of the far flow in Q2D, equation (\ref{flowq2d}), differs
from the far flow due to a local disturbance in a purely 2D
liquid. The latter is governed again by (2D) momentum diffusion,
leading to a logarithmic distance dependence. In 1D only a mass
monopole creates flow. Hence, in a Q1D geometry (confinement in a
linear channel) both the momentum and mass contributions are
exponentially small in $\tilde{r}/w$, leading to a short-ranged
hydrodynamic interaction \cite{PRL02}.

Exact calculations of the flow due to a point force (stokeslet), \ie
the Oseen tensors in the confined geometries, confirm the above
conclusions \cite{Mochon,ShaharBlake}.  Those exact results are
relevant to colloid motion in the limit of very small particles, $a\ll
w,\tilde{r}$. The strength of the heuristic arguments presented above is
their general validity---the conclusions concerning the far flow
should hold regardless of particle size or details of the confining
boundaries.  These arguments also imply that, for a spherical
particle, particle rotation does not contribute to the far flow under
confinement, since it perturbs only the liquid momentum.

When the suspension is very dilute we may consider the interaction
between two isolated particles. Suppose that the motion of particle 1
exerts on the liquid a unit force in the $x$ direction. The
disturbance will create a flow field which, in turn, will entrain
particle 2 located at $\vecr$. Thus, the coupling mobilities at large
distances are obtained as $\Delta_\rmL=v_x(\rho\xhat)$ and
$\Delta_\rmT=v_x(\rho\yhat)$, where $\vecv(\vecr)$ is given by
equation (\ref{flowq2d}) and $\xhat,\yhat$ denote unit vectors in the
$x$ and $y$ directions, respectively. This yields
\begin{equation}
  \phi\rightarrow 0:\ \ \ 
  \Delta_\rmL(\rho\gg 1) = \lambda/\rho^2,\ \ \ 
  \Delta_\rmT(\rho\gg 1) = -\lambda/\rho^2,
\label{Delta0}
\end{equation}
where $\lambda=f(z=0)$ is a coefficient which depends only on $a/w$.
(In the limit $a/w\ll 1$ one finds $\lambda=9/16$ \cite{Mochon}.)
As found in equation (\ref{Delta0}), the pair hydrodynamic interaction
in Q2D is very different from its unconfined counterpart. The decay
with distance is faster, $\sim 1/r^2$ instead of $\sim 1/r$, yet the
interaction is still long-ranged
\cite{Ajdari}. (Its decay is {\it slower} than near a single surface,
where the interaction falls off as $1/r^3$
\cite{1wall}.) The transverse coupling is negative, \ie particles
exert ``anti-drag'' on one another as they move perpendicular to their
connecting line. (In the unconfined case one has
$\Delta_\rmT=\Delta_\rmL/2$.) These properties are direct consequences
of the far flow field inferred above. The $1/\rho^2$ decay is that of
the flow due to a 2D mass dipole. The negative transverse coupling is
a result of the circulation flows in the dipolar field 
\cite{PRL04}.

\section{Three-body correction}
\label{sec_three}

As $\phi$ is increased, the pair hydrodynamic interaction should
become affected by the presence of other particles. If hydrodynamic
screening were to set in, we would expect the interaction at distances
much larger than the typical inter-particle distance to have the
form of equation (\ref{Delta0}) yet with a modified, $\phi$-dependent
prefactor. However, this is not the case, as is demonstrated below
both theoretically and experimentally.

We begin again with particle 1, located at the origin and exerting a
unit force in the $x$ direction. This creates a flow of the form
(\ref{flowq2d}), which entrains particle 2, located at $\brho$, with
velocity $\vecv(\brho)$. We now introduce particle 3 at $\brho'$. 
Particle 3 will obstruct the flow, thereby exerting an extra force
$\vecf_3$ on the liquid. This force is related to variations of the
flow $\vecv(\brho')$ over the volume now occupied by particle 3.  For
example, in an unconfined, isotropic system, according to Faxen's
first law \cite{colloids}, $\vecf_3\sim\nabla^2\vecv(\brho')$. In the
anisotropic Q2D geometry this law is modified to
\begin{equation}
  f_{3i} = C'_{ij}\nabla_\perp^2v_j(\brho',0) + 
  C''_{ij}\dzz|_{z=0}v_j(\brho',z),\ \ \ i,j=x,y.
\label{f3}
\end{equation}
[This result is readily obtained by inspecting all the ways to produce
a vector $\vecf_3$ from derivatives of another vector $\vecv$ while
using the various symmetries of the field (\ref{flowq2d}).] In
equation (\ref{f3}) $C',C''$ are $2\times 2$ coefficient tensors
(which can be found by a more detailed calculation),
and $\nabla_\perp^2\equiv\partial_{xx}+\partial_{yy}$. Since the flow
(\ref{flowq2d}) satisfies $\nabla_\perp^2\vecv=0$ and
$\dz^2|_{z=0}\vecv\sim\vecv$, equation (\ref{f3}) is simplified to
\begin{equation}
  f_{3i} = C_{ij}v_j(\brho').
\end{equation}
This local relation between $\vecf_3$ and $\vecv$ must be invariant to
rotation in the $(x,y)$ plane, which leads to $C_{xx}=C_{yy}\equiv
C_\rmL$ and $C_{xy}=-C_{yx}\equiv C_\rmT$.  Since the problem is
linear, the effect of the extra force on the velocity at $\brho$
is found by projecting the same flow field,
equation (\ref{flowq2d}), from $\brho'$ onto $\brho$, $\delta v_x =
v_i(\brho-\brho')f_{3i}(\brho')$. 
We then get
\begin{eqnarray}
  \delta v_x(\brho,\brho') = && 
  C_\rmL[v_x(\brho-\brho')v_x(\brho')+v_y(\brho-\brho')v_y(\brho')] +
\nonumber\\
  && C_\rmT[v_x(\brho-\brho')v_y(\brho')-v_y(\brho-\brho')v_x(\brho')].
\label{delta1}
\end{eqnarray}

Equation (\ref{delta1}) presents the correction to the $x$-velocity of
the liquid at $\brho$ given a fixed position $\brho'$ of the third
particle. We are interested, however, in the correction averaged over
all possible positions $\brho'$. This requires the probability density
$p(\rho')$ of finding a particle a distance $\rho'$ away from particle
1, which we assume here to be uniform, $p = \phi/(\pi a^2)$.  We
average according to $\langle\delta v_x\rangle = \int d^2\rho'p\delta
v_x(\brho,\brho')$ and find
\begin{equation}
  \langle\delta v_x\rangle(\brho) = \epsilon\phi\int d^2\rho'
  [v_x(\brho')v_x(\brho-\brho')+v_y(\brho')v_y(\brho-\brho')].
\label{delta2}
\end{equation}
The terms proportional to $C_\rmT$ cancel, leaving us with only one
coefficient, $\epsilon=C_\rmL/(\pi a^2)$.

For $\rho'\gg\rho$ the integrand in equation (\ref{delta2}) decreases
as $(\rho')^{-4}$. Hence, the integrated contributions from far-away
third particles should lead to a correction proportional to
$\phi/\rho^2$ which, as expected, would renormalise the prefactor of
the leading $1/\rho^2$ term.  However, carefully carrying out the
convolution in equation (\ref{delta2}), as shown in the Appendix,
proves this conclusion wrong. The specific angular dependence of the
flow (\ref{flowq2d}) leads to cancellation of terms, and the integral
of equation (\ref{delta2}) {\em vanishes}. Thus, the obstructions from
numerous particles add up to nothing, and the long-range response of
the suspension is identical to that of the particle-free liquid (to
linear order in $\phi$). Equation (\ref{Delta0}), therefore, remains
unchanged for low but finite concentrations. Deviations of the flow
from the far-field dipolar form and the introduction of a nonuniform
particle distribution (\ie static pair correlations) lead to
non-vanishing three-body effects. These corrections, however, are
short-ranged and thus do not change the above result. (The
short-ranged three-body effects will be addressed in a separate
publication.)

\section{Experimental results}
\label{sec_exp}

The experimental system consists of an aqueous suspension of
monodisperse silica spheres (diameter $2a=1.58\pm 0.04$ $\mu$m,
density 2.2 g/cm$^3$, Duke Scientific), tightly confined between two
parallel glass plates in a sealed thin cell (figure \ref{fig_system}).
The inter-plate separation is $w=1.76\pm 0.05$ $\mu$m, \ie slightly
larger than the sphere diameter, $2a/w\simeq 0.90$.  
Digital video microscopy and subsequent data analysis are
used to locate the centres of the spheres in the field of view and
then extract time-dependent two-dimensional trajectories.  Details of
the setup and measurement methods can be found elsewhere \cite{JCP01}.
Measurements were made at four values of area fraction, $\phi=$ 0.254,
0.338, 0.547, $0.619\pm 0.001$. From equilibrium studies of this
system \cite{JCP02} we infer that, for the purpose of this study, the
particles can be regarded as hard spheres.

The correlated Brownian motion of a particle pair is characterized by
a longitudinal and transverse coupling diffusion coefficients. Upon
scaling by $D_0a/w$, where $D_0=k_\rmB T/(6\pi\eta a)$ is the
Stokes-Einstein diffusion coefficient of a single unconfined sphere,
the two coefficients (thanks to the Einstein relation) become
identical to the two mobilities $\Delta_\rmL$ and $\Delta_\rmT$. The
two coupling mobilities are thus directly extracted from measured
particle trajectories as 
$
  \Delta_\rmL(\rho) = \langle x_1(t)x_2(t)\rangle_\rho/(2D_0t)
$ and 
$ \Delta_\rmT(\rho) = \langle y_1(t)y_2(t)\rangle_\rho/(2D_0t)
$,
where $x_i(t)$ and $y_i(t)$ are the displacements of particle $i$ of
the pair during a time interval $t$ along and transverse to their
connecting line, respectively.  The average
$\langle\rangle_\rho$ is taken over all pairs whose mutual
distance falls in a narrow range ($\pm 0.09$ $\mu$m) around
$\tilde{\rho}=w\rho$.

\begin{figure}[t]
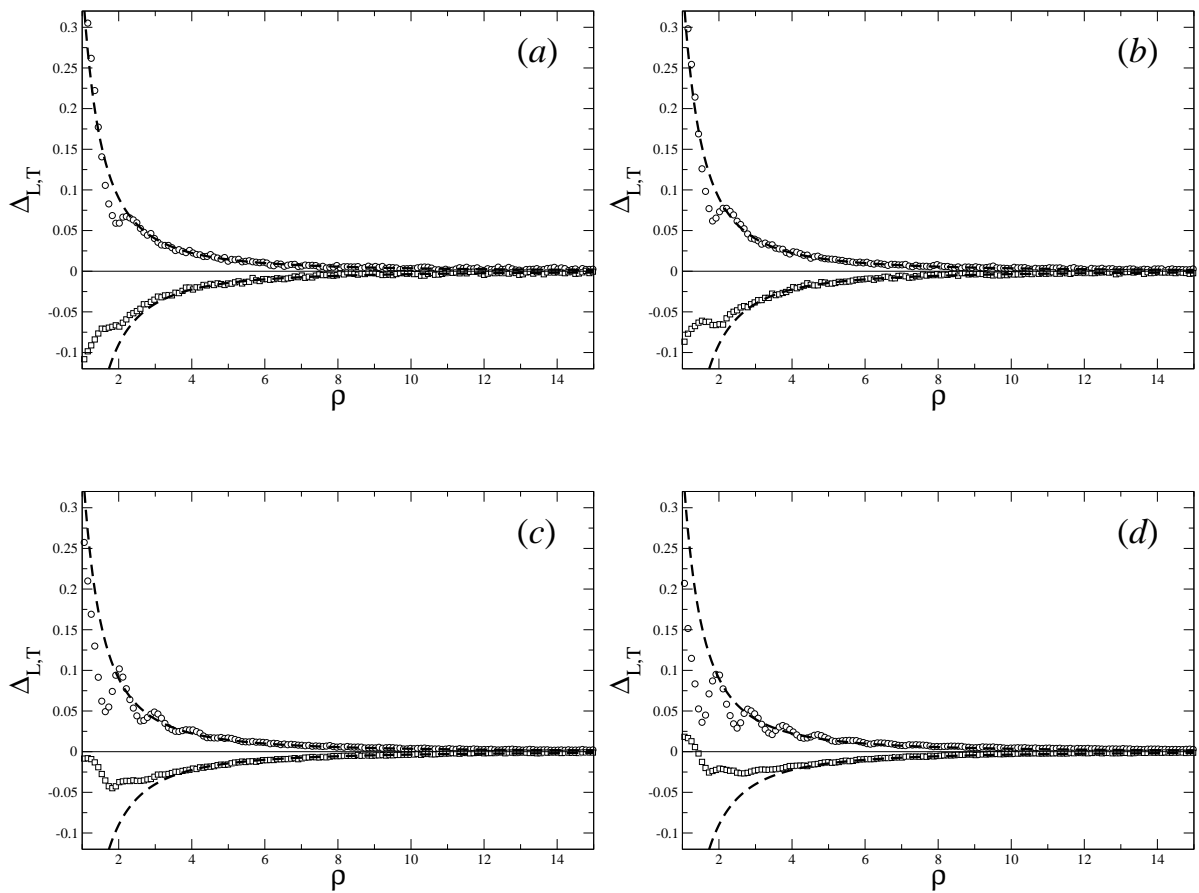

\vspace{0.7cm}
\centerline{\resizebox{0.5\textwidth}{!}
{\includegraphics{fig2a.eps}}
\resizebox{0.5\textwidth}{!}
{\includegraphics{fig2b.eps}}}
\vspace{1cm}
\centerline{\resizebox{0.5\textwidth}{!}
{\includegraphics{fig2c.eps}}
\resizebox{0.5\textwidth}{!}
{\includegraphics{fig2d.eps}}}
\caption[]{Longitudinal ($\Delta_\rmL$, circles) and transverse
($\Delta_\rmT$, squares) coupling diffusion coefficients as a function
of inter-particle distance $\rho$. The coefficients are scaled by
$D_0a/w$ and the distance by $w$. Area fractions are $\phi=0.254$
($a$), 0.338 ($b$), 0.547 ($c$), and 0.619 ($d$). Dashed lines are a
fit to $\pm\lambda/\rho^2$ with the same value of
$\lambda=0.36$ for all panels.}
\label{fig_exp}
\end{figure}

The measured coupling diffusion coefficients are presented in figure
\ref{fig_exp}. The leading behaviour for all $\phi$ values fits well
the predicted $\pm\lambda/\rho^2$ dependence of equation
(\ref{Delta0}).  The negative sign of $\Delta_\rmT$ confirms the
predicted ``anti-drag'' between particles located transverse to the
direction of motion. The fact that the fitted value of $\lambda=0.36$
does not change with $\phi$ demonstrates the absence of hydrodynamic
screening up to an area fraction of $0.619$. (Larger area fractions
could not be checked because the suspension began to crystallise
\cite{JCP02}.)

\section{Discussion}
\label{sec_dis}

Flows due to local disturbances in a Q2D geometry are governed by mass
propagation rather than momentum diffusion. We have demonstrated the
strong effect that this has on the pair hydrodynamic interaction
between colloid particles. The $1/r^2$ decay of the coupling with
distance is faster than in the unconfined case but slower than near a
single surface. The transverse coupling becomes negative. Three-body
effects do not renormalise the coupling coefficient, in contrast with
the usual case of hydrodynamic screening. Since momentum diffusion
does not contribute to the large-distance coupling, it is perhaps
natural that the flow effects of numerous distant particles may not
renormalise the viscosity, which is the transport coefficient
associated with momentum diffusion.

We have not treated the effect of particle motion perpendicular to the
boundaries.  Such fluctuations must exist in practice, yet our
experimental results suggest that they have a minor effect on the
long-distance behaviour. This is expected since the flows produced by
transverse fluctuations decay exponentially with distance
\cite{Mochon}.

We have restricted the discussion to far-field effects. At small
inter-particle distances various features are observed in the
hydrodynamic interactions (see Fig.\ \ref{fig_exp}), which are related
to the static particle pair correlation. We shall address these
short-ranged effects in a forthcoming publication.

\ack

We thank Michael Cates, Shigeyuki Komura and Tom Witten for helpful
discussions.  This research was supported by the Israel Science
Foundation (77/03), the National Science Foundation (CTS-021774 and
CHE-9977841) and the NSF-funded MRSEC at The University of Chicago.
H.D.\ acknowledges additional support from the Israeli Council of
Higher Education (Alon Fellowship).

\section*{Appendix}

We wish to calculate the convolution integral appearing in equation
(\ref{delta2}):
\begin{equation}
  I(\brho) = \int d^2\rho' [v_x(\brho')v_x(\brho-\brho')
  + v_y(\brho')v_y(\brho-\brho')],
\end{equation}
where (omitting the prefactor)
\begin{equation}
  v_x(\brho)=(x^2-y^2)/\rho^4,\ \ \ 
  v_y(\brho)=2xy/\rho^4.
\label{farflow}
\end{equation}
Calculating $I(\brho)$ by direct integration is quite tricky for
reasons similar to those encountered in electrostatics. Excluding two
small areas $\sim a^2$ around the (integrable) singularities at
$\rho'=0$ and $\brho'=\brho$, we divide the integration into three
domains ($\varphi$ and $\varphi'$ being the polar angles of $\brho$
and $\brho'$, respectively): (i) $a<\rho'<\rho-a$,
$0\leq\varphi'<2\pi$; (ii) $\rho'>\rho+a$, $0\leq\varphi'<2\pi$; (iii)
$\rho-a<\rho'<\rho+a$, $\varphi+a/\rho<\varphi'<\varphi+2\pi-a/\rho$.
The contribution from the inner domain (i) vanishes upon angular
integration for any $a$. The outer domain (ii) contributes $\pi/\rho^2
+ O(a/\rho^3)$. However, integration over the intermediate narrow
domain (iii) just cancels the outer contribution, yielding
$-\pi/\rho^2 + O(a/\rho^3)$. Thus, in the limit $a\rightarrow 0$ the
result is $I=0$. A finite value of $a$ leads to a correction of
$O(a/\rho^3)$, which is negligible compared to the leading $1/\rho^2$
coupling at large enough distances.

The direct integration pitfalls can be circumvented altogether by
switching to Fourier space, $\cF[f(\brho)]\equiv f(\vecq)\equiv\int
d^2\rho e^{-i\vecq\cdot\brho}f(\brho)$. The transformed velocity field
is
\begin{equation}
  v_x(\vecq) = -\pi(q_x^2-q_y^2)/q^2,\ \ \ 
  v_y(\vecq) = -\pi(2q_xq_y)/q^2.
\label{cFDelta}
\end{equation}
Using equation (\ref{cFDelta}) in the convolution readily gives
\begin{equation}
  I(\brho) = \pi^2 \cF^{-1}[1] = \pi^2 \delta(\brho).
\end{equation}
Hence, $I=0$ for any $\rho>0$.

\section*{References}


\begin{thebibliography}{99}

\bibitem{colloids}
Russel W B, Saville D A and Schowalter W R 1989
{\it Colloidal Dispersions} (New York: Cambridge University Press)

\bibitem{DoiEdwards}
Doi M and Edwards S F 1986
{\it The Theory of Polymer Dynamics}
(New York: Oxford University Press)

\bibitem{1wall}
Perkins G S and Jones R B 1992
{\it Physica A} {\bf 189} 447
\nonum
Dufresne E R, Squires T M, Brenner M P and Grier D G 2000
{\it Phys.\ Rev.\ Lett.} {\bf 85} 3317 

\bibitem{2wall}
Lobry L and Ostrowsky N 1996
{\it Phys.\ Rev.\ B} {\bf 53} 12050
\nonum
Lin B, Yu J and Rice S A 2000
{\it Phys.\ Rev.\ E} {\bf 62} 3909
\nonum
Dufresne E R, Altman D and Grier D G 2001
{\it Europhys.\ Lett.} {\bf 53} 264

\bibitem{container}
Segre P N, Herbolzheimer E and Chaikin P M 1997
{\it Phys.\ Rev.\ Lett.} {\bf 79} 2574

\bibitem{PRL02}
Cui B, Diamant H and Lin B 2002
{\it Phys.\ Rev.\ Lett.} {\bf 89} 188302

\bibitem{PRL04} 
Cui B, Diamant H, Lin B and Rice S A 2004
{\it Phys.\ Rev.\ Lett.} {\bf 92} 258301

\bibitem{Nagele}
Pesche R and Nagele G 2000
{\it Europhys.\ Lett.} {\bf 51} 584
\nonum
Pesche R and Nagele G 2000
{\it Phys.\ Rev.\ E} {\bf 62} 5432

\bibitem{Zangi}
Zangi R and Rice S A 2004
{\it J.\ Phys.\ Chem.\ B} {\bf 108} 6856

\bibitem{Finland}
Chvoj Z, Lahtinen J M and Ala-Nissila T 2004
{\it J.\ Stat.\ Mech.\ Theor.\ Exp.} P11005
\nonum
Falck E, Lahtinen J M, Vattulainen I and Ala-Nissila T 2004
{\it Eur.\ Phys.\ J.\ E} {\bf 13} 267

\bibitem{Happel}
Happel J and Brenner H 1965
{\it Low Reynolds Number Hydrodynamics}
(Englewood Cliffs: Prentice-Hall)

\bibitem{Mochon}
Liron N and Mochon S 1976
{\it J.\ Eng.\ Math.} {\bf 10} 287

\bibitem{ShaharBlake}
Liron N and Shahar R 1978
{\it J.\ Fluid Mech.} {\bf 86} 727
\nonum
Blake J R 1979
{\it J.\ Fluid Mech.} {\bf 95} 209

\bibitem{Ajdari}
Long D, Stone H A and Ajdari A 1999
{\it J.\ Colloid Interface Sci.} {\bf 212} 338
\nonum
Long D and Ajdari A 2001
{\it Eur.\ Phys.\ J.\ E} {\bf 4} 29

\bibitem{JCP01}
Cui B, Lin B and Rice S A 2001
{\it J.\ Chem.\ Phys.} {\bf 114} 9142

\bibitem{JCP02}
Cui B, Lin B Sharma S and Rice S A 2002
{\it J.\ Chem.\ Phys.} {\bf 116} 3119

\end{thebibliography}
\end{document}